# THE NEW SET-UP IN THE BELGRADE LOW-LEVEL AND COSMIC-RAY LABORATORY

by

*Aleksandar* DRAGIĆ[1*], *Vladimir I.* UDOVIČIĆ[1], *Radomir* BANJANAC[1],
*Dejan* JOKOVIĆ[1], *Dimitrije* MALETIĆ[1], *Nikola* VESELINOVIĆ[1], *Mihailo* SAVIĆ[2],
*Jovan* PUZOVIĆ[2], *and Ivan V.* ANIČIN[1]

[1]Institute of Physics, Belgrade, Serbia
[2]Faculty of Physics, University of Belgrade, Belgrade, Serbia

The Belgrade underground laboratory consists of two interconnected spaces, a ground level laboratory and a shallow underground one, at 25 meters of water equivalent. The laboratory hosts a low-background gamma spectroscopy system and cosmic-ray muon detectors. With the recently adopted digital data acquisition system it is possible to simultaneously study independent operations of the two detector systems, as well as processes induced by cosmic-ray muons in germanium spectrometers. Characteristics and potentials of the present experimental setup, together with some preliminary results for the flux of fast neutrons and stopped muons, are reported here.

## INTRODUCTION

The low-level and cosmic-ray laboratory in Belgrade is dedicated to the measurement of low activities and cosmic-ray (CR) muon components. At the intersection of the two research subjects, the study of muon-induced background in gamma spectroscopy is of particular interest. The laboratory adds to the list of relatively shallow underground laboratories worldwide (see the recent review [1]). It is located on the right bank of the river Danube in the Belgrade borough of Zemun, on the grounds of the Institute of Physics. The ground level portion of the laboratory (GLL), at 75 meters above sea level (m.a.s.l), is situated at the foot of a vertical loess cliff, about 10 meters high. The underground part of the laboratory (UL), useful area 45 m$^2$, is dug into the foot of the cliff and is accessible from the GLL via a 10 meters long horizontal corridor which also serves as a pressure buffer for a slight overpressure that is constantly maintained in the UL (fig. 1). The overburden of the UL is about 12 meters of loess soil, equivalent to 25 meters of water. The container, which is to accommodate the top laboratory (TL), is situated at the top of the cliff, just above the UL. The GLL and UL have been in some use for a number of years now, while the TL is still not functional.

Continuous measurements of the cosmic-ray muon flux by means of a pair of small plastic scintillators 50 cm × 23 cm × 5 cm started in the GLL and UL back in 2002 and lasted for about 5 years. These measurements yielded the precise values of the integral CR muon flux at ground level and at the depth of 25 m.w.e. [2]. Different analyses of the time series of these measurements have also been performed [3, 4].

Significant efforts are being made to contain the low radon concentration within the laboratory. The UL is completely lined with a hermetically sealed, 1 mm thick aluminum foil. The ventilation system maintains the overpressure of 2 mbar, so as to prevent radon diffusion from the soil. Fresh air entering the laboratory is passed through a two-stage filtering system. The first stage is a mechanical filter for dust removal. The second one is a battery of coarse and fine charcoal active filters. The concentration of radon is kept at an average value of about 10 Bq/m$^3$. Throughout the years, certain interesting behaviors of the said concentration have also been reported [5, 6].

The two laboratory spaces have recently been furnished with a new experimental set-up which is now ready for routine measurements. Here presented are some preliminary results of wider interest, ob-tained during the testing period of the new equipment and the development of the needed software.

---

* Corresponding author; e-mail: dragic@ipb.ac.rs

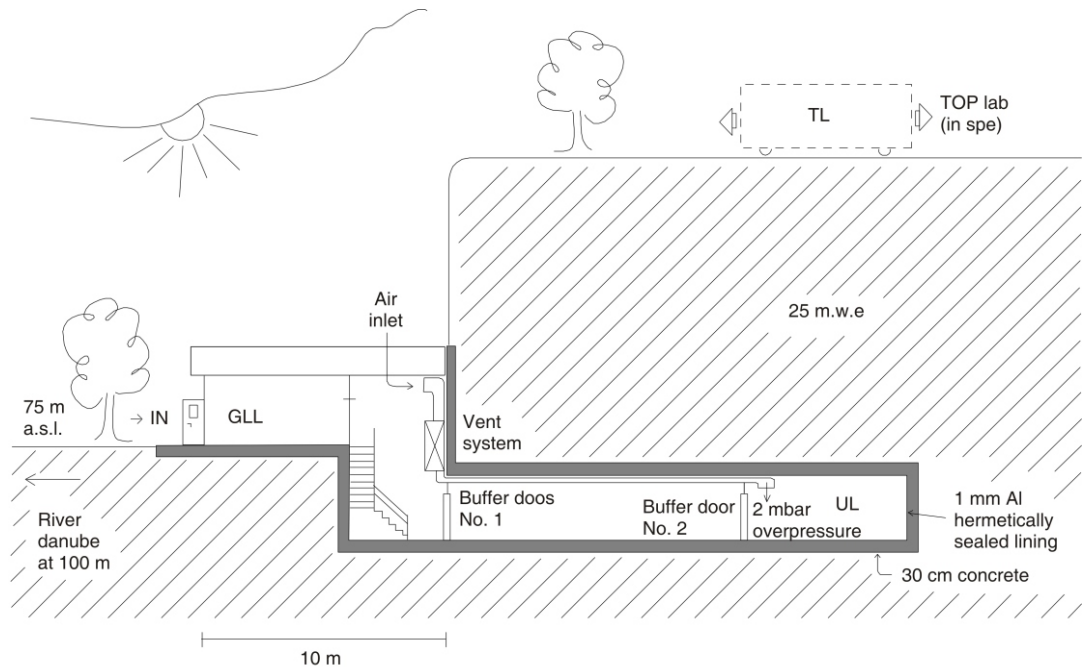

**Figure 1. Cross-section of the low-level and CR laboratory at IOP, Belgrade, 44°49'N, 20°28'E, vertical rigidity cutoff 5.3 GV**

**EXPERIMENTAL SET-UP**

The equipment now consists of two almost identical sets of detectors and analyzing electronics, one set situated in the GLL, the other in the UL. Each set is composed of muon detectors and a gamma spectrometer.

A pair of plastic scintillator detectors is used for CR muon measurements. One of them is a larger 100 cm 100 cm 5 cm detector, equipped with four PMT directly coupled to the corners beveled at 45°, made by Amcrys-H, Kharkov, Ukraine. The other, a small 50 cm 23 cm 5 cm plastic scintillator detector, with a single PMT looking at its longest side via a Perspex light guide tapering to the diameter of a PMT, made by JINR, Dubna, Russia, and assembled locally. The smaller detector may serve as a check of stability of the muon time series obtained from the larger detector, which is important for long term measurements. It can also be used (in coincidence with the larger detector) for measurements of the lateral spread of particles in CR showers. Plastic scintillation detectors are also employed for active shielding of gamma spectrometers.

In the UL, a 35% efficiency radio-pure p-type HPGe detector, made by ORTEC, in its 12 cm thick cylindrical lead castle, is deployed. Another HPGe detector, of 10% efficiency, is put to use in the GLL. It is shielded with lead of the same origin, parts of a plumbing system collected at a demolition site of an old housing estate. The exact history of this lead is not known, but all the components are known to be older than two half-lives of Pb-210.

At the heart of the data acquisition system are two flash: analog to digital converter (ADC), flash analog to digital converter (FADC), one in each laboratory, made by CAEN (type N1728B). These are versatile instruments, capable of working in the so-called energy histogram mode when performing as digital spectrometers or, in the oscillogram mode, when they perform as digital storage oscilloscopes. In both modes, they sample at 10 ns intervals into $2^{14}$ channels in four independent inputs. The full voltage range is 1.1 V.

They are capable of operating in the list mode, when every analyzed event is fully recorded by the time of its occurrence and its amplitude. This enables the correlation of events, both prompt and arbitrarily delayed, at all four inputs with the time resolution of 10 ns. Single and coincident data can be organized into time series within any integration period from 10 ns up. The two N1728B units are synchronized, enabling coincidence/correlation of the events recorded in both of them. The flexible software encompassing all above said off-line analyses is user-friendly and entirely homemade.

The usual disposition of FADC inputs is described next. The preamplifier outputs of the PMT of the larger detectors are paired diagonally, the entire detector thus engaging these two inputs of the FADC. Signals from these inputs are later coincided off-line and their amplitudes added to produce the single spectra of these detectors. This procedure results in a practically complete suppression of the uninteresting low-energy portion of the background spectrum (up to some 3 MeV), mostly due to environmental radiation, leaving only high-energy loss events due to CR muons and EM showers that peak at about 10 MeV. The output of the PMT of the smaller detector is fed to the third input. The fourth input is reserved for the HPGe detectors.

In some instances, auxiliary measurements are performed with a different definition of the inputs of the data acquisition system. For example, a (3 × 3)" NaI detector is used in the GLL to scan the response of the larger detector to CR as a function of the position of the interaction point.

In the UL, the HPGe detector is positioned beneath the center of the larger detector (fig. 2). For the purpose of measuring low activities, the large plastic detector is used in anticoincidence, as a cosmic-ray muon veto detector. In order to study the effects of cosmic rays on the spectra in low-level high-resolution gamma-ray spectroscopy, it is used in coincidence as the trigger for the CR-induced processes. These two functions of the system are performed simultaneously and do not interfere, as they are realized by different off-line analyses of the same set of data.

## TESTING THE SYSTEM AND DEVELOPING THE SOFTWARE

In order to test the performance of the digital spectroscopy system, a series of test measurements with different count rates and different types of radiation detectors at the input of the FADC are performed.

One of the tests is designed to correspond to the real situations where neutrons created by CR muons in the lead shield produce certain effects in HPGe detectors. Neutrons produced by muons in the vicinity of the detector or the surrounding rock mass represent a significant source of background in ultra-low background experiments carried out deep underground, such as those searching for dark matter or double beta decay. In the test, done at the GLL, Cf-252 was used as a neutron source and the small plastic scintillator as a trigger for neutrons. To distinguish between the effects of fast and slow neutrons, some materials common in neutron work, such as rubberized $B_4C$, Cd sheets, paraffin, lead and iron slabs, were placed around and in between the source and the detectors. In addition to the environmental background, the HPGe spectra consists of different features induced by slow and fast neutrons in the HPGe detector and surrounding materials.

Results of measurements are stored as a list of events represented with their amplitudes, time tags, designation of input channels and some additional information (pile-up event or not, *etc*.). The time tag for every event is determined by the moment of crossing the set-triggering level. In order to minimize the amplitude walk, there is a possibility to choose between different types of triggers, termed here as simple, digital or CFD. We have stuck to the digital trigger which was found to work reliably and, if necessary, to the off-line correction of the amplitude walk [7].

The distribution of time intervals between events in the trigger detector and the HPGe is deduced from the recorded data in off-line analysis. There is no need to implement the hardware tome-to-amplitude converter (TAC). For convenience, in what follows we will refer to this distribution as the TAC spectrum.

As an illustration, the TAC spectrum between events in the plastic scintillator and HPGe is presented here (fig. 3).The prompt-time distribution is seen to be about 90 ns wide, while the tail of delayed coincidences is discernable beyond approx. 100 ns upon the prompt peak. The same time spectrum, off-line corrected for the amplitude walk, according to the procedure described in [7], is presented in fig. 3(b). The full

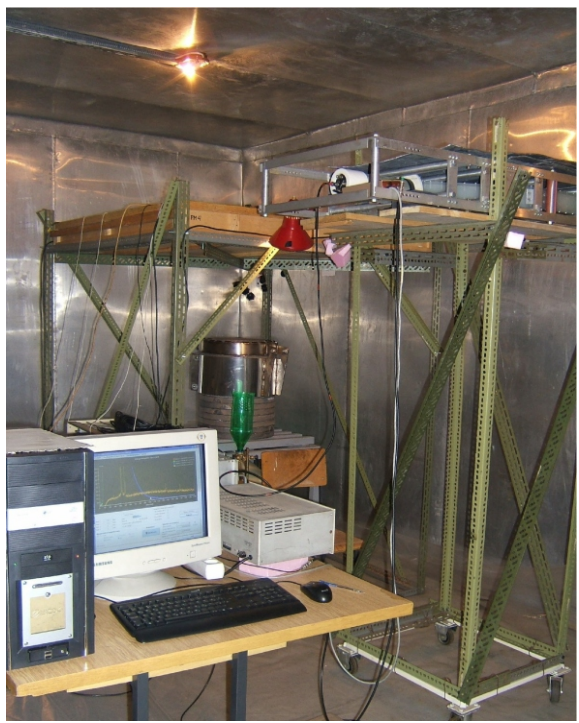

**Figure 2. Detectors in the underground laboratory (UL). The big plastic scintillator is positioned over the HPGe detector, seen in its lead shielding. The small plastic scintillator is in the front upper right corner. A hermetically sealed, 1 mm Al lining covering the entire UL, which enables the doubly filtrated ventilation system to sustain an overpressure of 2 mbar and keeps the radon concentration at an average level of some 10 Bq/m³, is also shown**

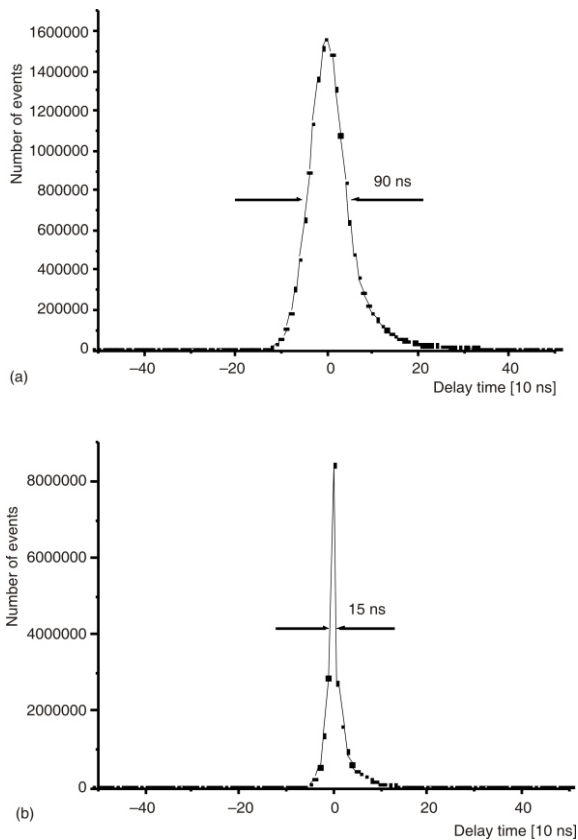

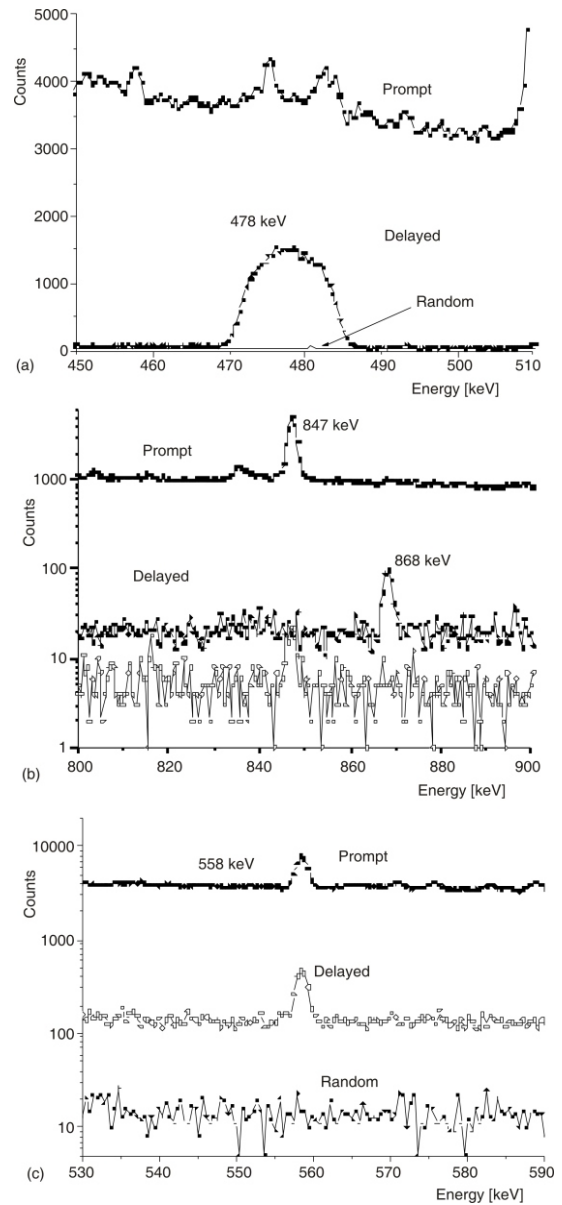

Figure 3. (a) The time spectrum between events in the plastic scintillator and the HPGe; (b) the same time spectrum, but with off-line time corrected for the amplitude walk

with at half maximum (FWHM) is now only 15 ns and delayed coincidences, perhaps, start as early as at approximately 20 ns after the prompt peak. Minding the geometry of our set-up, we expect the effects induced by fast neutrons, both in the environment and the HPGe detector itself, to be within the prompt peak, while those induced by thermalized neutrons should be found in the tail of delayed coincidences.

To illustrate the complete separation of the effects due to fast and slow neutrons, here presented (fig. 4) are the portions of the coincident HPGe spectrum around the spectral lines originating from different processes induced by neutrons in different materials, gated with different portions of the time spectrum – with the prompt peak, tail of delayed events up to one microsecond and the flat portion of random coincidences. Short comments can be found in the caption under fig. 4.

We will now briefly comment on the two well-known structures induced by neutrons in the HPGe detector itself, the structures at 596 keV and 692 keV. Their appearance in the coincidence spectra is depicted in fig. 5. The triangular distributions result from the summing of the radiations depopulating the state ex-cited in inelastic neutron scattering with the energy of the recoiling nucleus. The one at 596 keV appears in the prompt spectrum, since the state at 596 keV in Ge-74 is short-lived. The regular peak of this energy in the delayed spectrum results from the thermal neutron capture by Ge-7, as is the case with the neighboring 609 keV line stemming from the same capture reaction. If the neutron flux at the detector is high, some of the intensity of the ubiquitous background line of 609 keV, usually entirely attributed to $^{214}$Bi, is due to this process.

Figure 4. The prompt, delayed (up to 1 $\mu$s), and random coincidence spectra of: (a) the 478 keV Doppler-widened line from the (n, $\alpha$) reaction on $^{10}$B which appears only in the delayed spectrum; (b) the 847 keV line from the (n, n') inelastic scattering on $^{56}$Fe which appears only in the prompt spectrum; (c) the 558 keV line which appears in both the prompt and delayed spectra, proving that this line originates partly from the usually assumed thermal neutron capture by $^{113}$Cd and, depending on the hardness of the neutron spectrum, in part, from the fast neutron (n, n') reaction on $^{114}$Cd

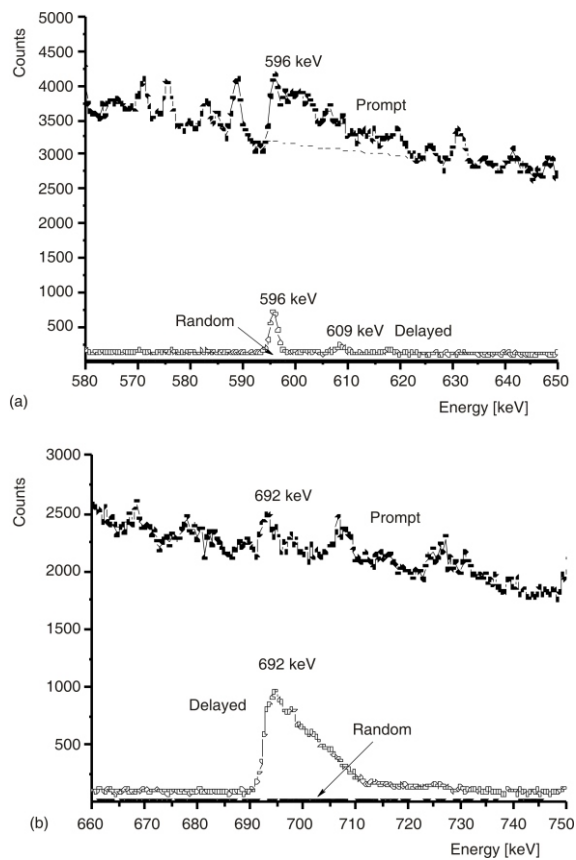

**Figure 5. The spectrum of prompt, delayed, and random coincidences (practically negligible) containing: (a) the structure at 596 keV from inelastic fast neutron scattering on Ge-74 in the prompt spectrum, and the line of the same energy from slow neutron capture on Ge-73, in the delayed spectrum; (b) the structure at 692 keV from the inelastic scattering of fast neutrons on Ge-72, this time in the delayed spectrum, due to the finite lifetime of 444 ns at the excited state of 692 keV**

The structure at 692 keV, however, appears in the delayed spectrum since the excited state of this energy in Ge-72 is comparatively long-lived, with a half-life of 444 ns, which is long in comparison to the time resolution of our system. As a demonstration of the capabilities of the system, we determined this half-life by setting the software gate to encompass the whole triangular structure in the coincident Ge spectrum, thus producing a time spectrum corresponding to this condition, thanks to which the fit produced a satisfactory value of 447(25) ns for the said half-life.

This particular structure has been studied in detail many times in the past, since 692 keV radiation is pure E0, detectable with 100% efficiency, which is why the integral of the triangular structure is a reliable measure of the fast neutron flux at the position of the detector [8-12]. These studies were performed with analog spectroscopy systems where the integration constants are long and the recoils invariably sum up with the 692 keV pulses. In digital spectroscopy systems, however, there is one important caveat to keep in mind when using the integral of this structure for fast neutron flux determination. It appears that here the shape and the intensity of the distribution strongly depend on the height of the triggering level. The recoil pulse is prompt, while the corresponding 692 keV pulse follows the recoil with delay distributed according to the decay law with the half-life of 444 ns. When the trigger is higher than the height of the recoil pulse, 692 keV pulse sums practically completely with the recoil. When the trigger is lower than the recoil, it will trigger the ADC, and this pulse, together with the following 692 keV pulse, will be rejected by the pile-up rejecting algorithm. This is illustrated in fig. 6 where the same portion of the direct HPGe spectrum is presented, with two different triggering levels. The width of the triangular structure appears proportional to the height of the triggering level. If this structure is to be used for quantitative purposes, the safe height of the triggering level that may be recommended is, thus, perhaps about 50 keV, when the loss of the counts within the structure is expected to be negligible.

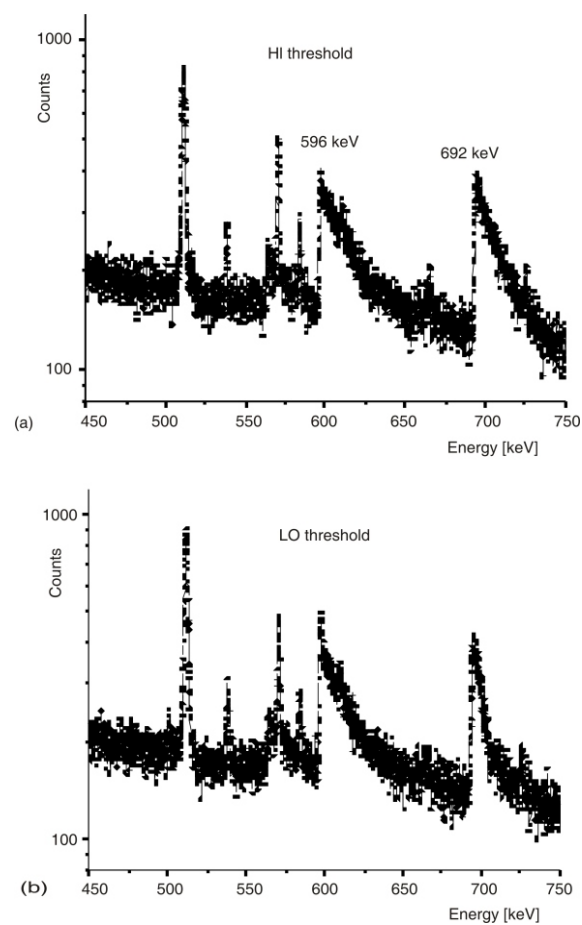

**Figure 6. The 692 keV distribution in the direct spectrum with the triggering level set at 50 keV (a) and at 20 keV (b)**

After the system satisfactorily passed all the tests, we started some preliminary measurements of the type that we plan to perform in the long run in the future, and it is of these results that we report in what follows.

**PLANNED MEASUREMENTS**

We plan to continuously collect the background spectra of all detectors in both laboratories, eventually changing only their mutual spatial arrangements. The spectra of the large scintillators will stretch up to a couple of hundreds of MeV, so as to include all multiple CR and shower events, while the one corresponding to the HPGe detector will go up to about 30 MeV, in order to include all possible nuclear radiations induced by CR radiations. Each event is to be recorded in a separate list, in accordance with the time of the occurrence of its trigger (with the resolution of 10 ns) and by its amplitude (in 32 k channels). By off-line analyses of this data we expect to directly obtain:

– the continuous time series of cosmic-ray intensity (muon plus electromagnetic – EM, components) in large and small plastic scintillators at ground level, as well as those generated underground,
– decoherence curves of cosmic-ray coincidence counts and coincidence spectra at different separations of the said detectors, be it at ground level or the underground one, the idea being to first define and, afterwards, separate muon from EM components,
– the spectrum of the HPGe detector in coincidence and anticoincidence with the large plastic scintillator positioned right above it, in the underground, and
– as well as the signatures of the soft component of EM showers in the spectrum of the unshielded NaI detector, taken in coincidence with the plastic detectors.

Since all above mentioned measurements are spectral, we hope to exploit this feature to some advantage, even in the case of rather featureless spectra of plastic scintillators. With the help of MC simulation programs (mainly CORSIKA and GEANT4), we expect to discriminate the signatures of CR muons from those of electromagnetic showers to some degree.

If all the measurements are performed continuously, we estimate that, together, both set-ups will produce about 1 TB of data per year, all of which would be kept permanently for later analyses.

To illustrate the potential of these measurements, we will now briefly report on some preliminary results obtained during a testing period, approximately yearlong. We will first briefly discuss the performance of low-level measurements and then those pertaining to CR measurements.

**LOW-LEVEL MEASUREMENTS**

Future applications of the low-background gamma spectroscopic system include the study of rare nuclear processes, measurements of environmental radioactivity and radiopurity of materials.

The cylindrical lead shielding of the 35% efficiency radio-pure HPGe ORTEC detector, with a wall thickness of 12 cm and an overall weight of 900 kg, was cast locally out of scratch plumbing retrieved after the demolition of some old housing. The integral background rate in the region from 50 keV to 3 MeV is about 0.5 cps. The lines of Co-60 are absent in the background spectrum, while the line of Cs-137 with the rate of $1 \cdot 10^{-4}$ cps starts to appear significantly only if the measurement time approaches one month. Fukushima activities, though strongly present in our inlet air filter samples, did not enter the background at observable levels, in spite of the great quantities of air that we pump into the UL to maintain the overpressure, and it seems that the double air filtering and double buffer door system, along with stringent radiation hygiene measures, is capable of keeping the UL clean in cases of global accidental contaminations (see *e. g.* [13]).

No signatures of environmental neutrons, neither slow nor fast, are present in direct background spectra. The rates of some characteristic background

**Table 1. Count rates in some background lines. Rotes are given in counts per second (cps)**

| Energy [keV] | Count rate [cps] |
|---|---|
| 186.2 (Ra-226) | $2.4 \cdot 10^{-4}$ |
| 351.9 (Pb-214) | $1.1 \cdot 10^{-3}$ |
| 583.1 (Tl-208) | $6.6 \cdot 10^{-4}$ |
| 609.3 (Bi-214) | $1.1 \cdot 10^{-3}$ |
| 911.1 (Ac-228) | $4.5 \cdot 10^{-4}$ |
| 1460.8 (K-40) | $3.5 \cdot 10^{-3}$ |
| 2614.5 (Tl-208) | $1.1 \cdot 10^{-3}$ |

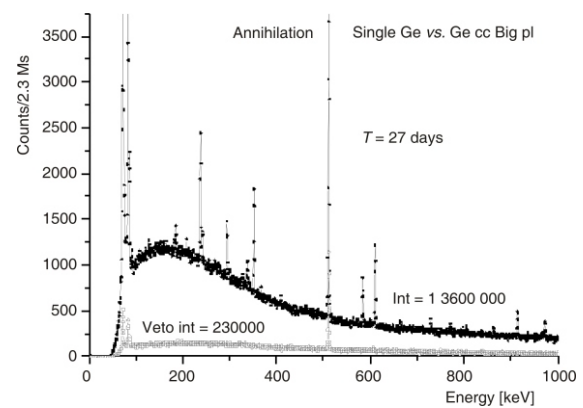

**Figure 7. The background spectrum of the HPGe detector in its lead castle in the UL and the part of the spectrum coincident with the large scintillator positioned right above it**

lines are listed in tab. 1. With the large plastic scintillator currently positioned rather high over the detector top, at a vertical distance of 60 cm from the top of the lead castle, in order to allow for the placing of voluminous sources in front of the vertically oriented detector, the off-line reduction of this integral count by the CR veto condition is about 18% (see fig. 7). Up to a factor of two might be gained if the veto detector were to be positioned at the closest possible distance over the HPGe detector. This agrees well with simple estimates of the rate of events susceptible to the veto condition [14]. The veto spectrum contains all events, prompt as well as those with delays of up to 10 s, which is why besides the continuum it contains only the lead X-rays and the annihilation line. As we shall see, the selection of delayed events only reveals some other details in this spectrum. Since for the time being we are not able to improve on the intrinsic background of our detector, when analyzing the analytical powers of our system, at present, we do not insist on the lowering of statistical errors which depend on background levels solely and are difficult to reduce further with available means, but rather emphasize its stability due to the low and controlled radon concentration in the laboratory. This is essential, especially in NORM measurements, and makes our system virtually free of systematic errors as compared to systems which operate in environments where radon is not controlled and where the reduction of post-radon background activities is achieved by flushing the detector cavity with liquid nitrogen vapor, where the transient regimes during sample changes and possible deposition of radon progenies [15] may introduce systematic uncertainties which are difficult to estimate.

**COSMIC-RAY MEASUREMENTS**

**Muon spectra and the time series**

During the commissioning of the large plastic detectors, we tested the response of these detectors to CR muons and their stability over a prolonged period of time. Certain results of these preliminary studies are presented here.

In fig. 8, we present the spectra of the two diagonals of the large plastic scintillator in the UL. Contrary to the situation in the GLL, the peak of charged particle energy losses situated at about 10 MeV (due to both muons and electrons from EM showers), is not fully separated from the low-energy tail of Compton electrons in the UL, because of gamma-ray interactions (both environmental and from EM showers).

Figure 9 presents the coincident sum spectra of the two diagonals. Energy spectra now contain only the well-defined peak of charged particles energy losses. The offsets are not imposed and occur simply

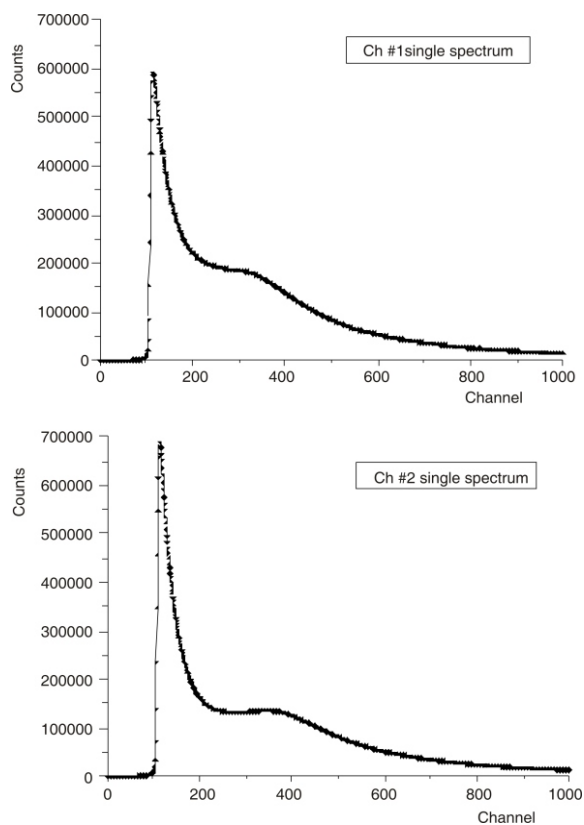

Figure 8. The spectra of two diagonals of the large plastic detector in the UL. Note that the peak of charged particle energy losses of 10 MeV, which corresponds to channel 320, is not separated from the low-energy tail of Compton scattered environmental gamma radiations. When summed, in coincidence they produce the spectrum from fig. 9

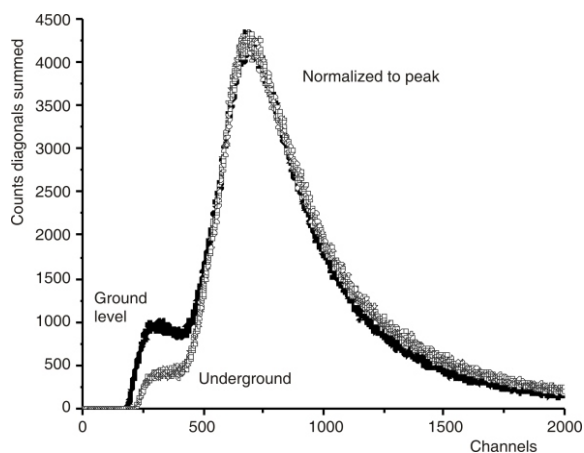

Figure 9. The sum spectra of two diagonals of the large plastic detectors in the UL and GLL. For comparison, the spectra are normalized for the peaks to coincide. Channel 650 now corresponds to the muon energy loss of 10 MeV. The integral of this peaked distribution is taken as the first approximation to the CR muon count by the large detectors

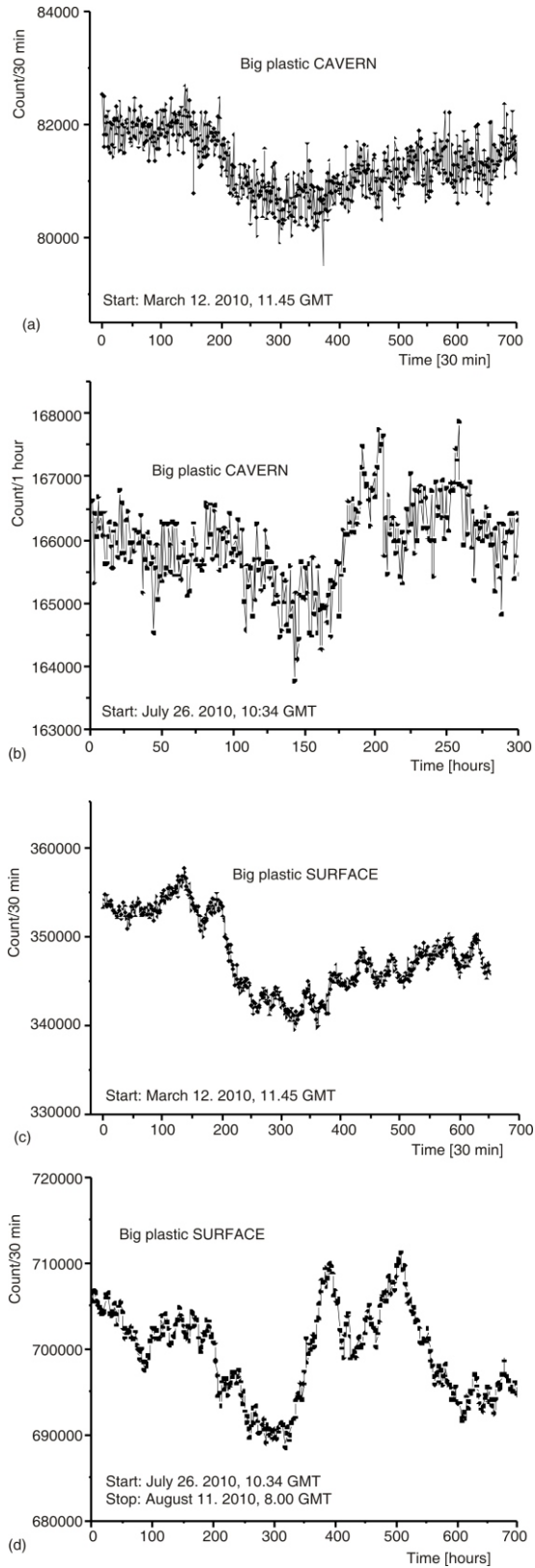

Figure 10. The time series of the CR muon count of the large plastic detector in the UL – (a) and (b) graphs – and GLL (c) and (d) graphs for the period starting with March 12, 2010, averaged to half hour intervals, as opposed to the period starting with July 26, 2010, averaging to one hour intervals. It is evident that the modulation in the two laboratories is correlated and that the amplitude of the modulation in the UL is roughly half

because at low energies there are no coincident events. As the simulations demonstrate, this is so because single Compton electrons do not produce enough light to trigger both diagonals.

The majority of events that produce this peaked distribution are due to CR muons that pass through the detector. We thus form the time series of this spectrum integrated over different time intervals. As an example, in fig. 10 we present the time series of this count in 30-minute intervals, both in the UL and in the GLL, for a period of 16 days in March 2010, and in one-hour intervals for the period starting with July 26, 2010. The data are not corrected either for atmospheric pressure or temperature.

The two series appear highly correlated, the amplitude of the modulation of this count in the UL is about 1.8%, while the corresponding one in the GLL is about 3.5%. At these integrating times, this is already sufficiently statistically significant, even in the UL.

Previous measurements at the same location with the small detectors yielded results for the muon flux of $1.37(6) \cdot 10^{-2}$ per $cm^2 s$ in the GLL, and of $4.5(2) \cdot 10^{-3}$ per $cm^2 s$ in the UL [2].

## NEUTRONS AND STOPPED POSITIVE MUONS IN THE UNDERGROUND LABORATORY

During the testing period, we have accumulated some six months of data-taking in the underground laboratory. The background spectrum of the HPGe de-tector containing the coincidences with the large plastic scintillator from the delayed tail of the corresponding TAC distribution in the region of 1 to 5 μs, shows at this statistics only two interesting features (fig. 11), though some more seem to emerge, but still insufficiently significant. The first is the already discussed triangular structure at 692 keV already discussed, which originates from the inelastic (n, n') scattering exciting the first excited state of the stable Ge-72 within the Ge detector itself. In this case the trigger was sufficiently high, and according to our finding, the intensity of the 692 keV distribution can be reliably used for the estimate of the fast CR-induced neutron flux at the position of the detector. To verify this, we applied the software gate to this structure and obtained the TAC distribution presented in fig. 12(a). Although the statistics is poor, the fit through the tail of delayed coincidences yields the half-life of 500(50) ns, which compares well with the known value of 444 ns. Using the expression from ref. [8], we obtain the value of 4(1) $10^{-7}$ cm$^{-2}$s$^{-1}$ for the flux of neutrons of CR origin with energies over 1 MeV. This refers to the flux at the depth of 25 m.w.e., (see *e. g.* [16]) within roughly a ton of lead, a common environment in most measurements of low activities.

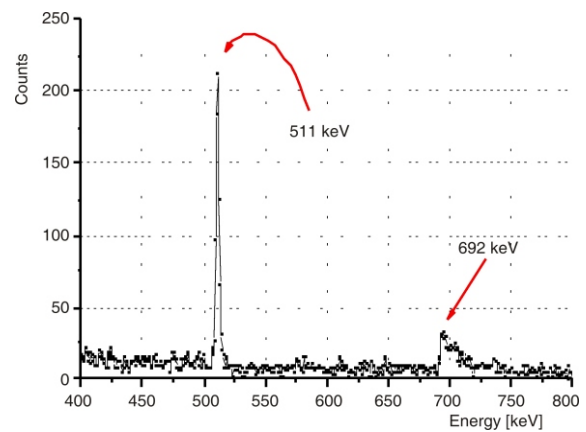

Figure 11. The portion of the background of the HPGe spectrum coincident with the large plastic detector with delays in the range of 1 to 5 $\mu$s, after 187 days of measurement time. It shows the annihilation line which is due to the decays of positive muons stopped in the lead castle, and the triangular structure at 692 keV, which is due to inelastic scattering of fast neutrons on 72-Ge, the neutrons originating mostly from direct fast muon interactions with nuclei and certainly less from captures of stopped negative muons. The threshold in this spectrum is sufficiently high to leave this last structure unscathed

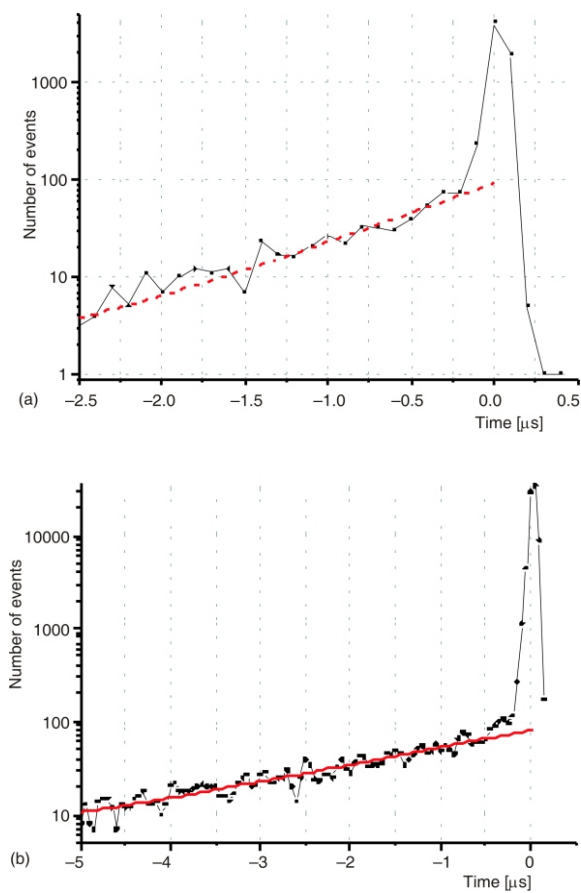

**Figure 12. Time distributions of events that belong (a) to the structure of 692 keV, the slope of which yields 500(50) ns for the half-life of the state of this energy in 72-Ge and, (b) that of the annihilation line, which yields 2.24(9) μs for the mean life of the muon**

The second feature of this spectrum is the annihilation line. The gate put on this line gives the TAC distribution presented in fig. 12(b), where the fit through the tail of delayed coincidences yields the mean life of 2.24(9) μs. This justifies the assumption that these events are due to the decays of stopped positive muons. We further assume that the source of these delayed annihilations is homogeneously distributed throughout the volume of the lead castle and use GEANT4 to find the overall detection efficiency. From the intensity of these delayed annihilations, we then obtain that the number of stopped positive muons per kg of lead per second equals 3.0(5) $10^{-4}$ $\mu_{stop}$/kgs.

We have also been able to estimate the number of stopped muons in the large plastic scintillators themselves, both in the GLL and the UL (*e. g.* [16]). For this purpose, we looked into the distribution of time intervals between the successive counts of these detectors. The gross structure of this distribution is nicely exponential, corresponding to the average CR counting rate and to an average time interval between the counts, reciprocal to the said rate. At short time intervals, however, the distribution strongly departs from this shape (fig. 13).

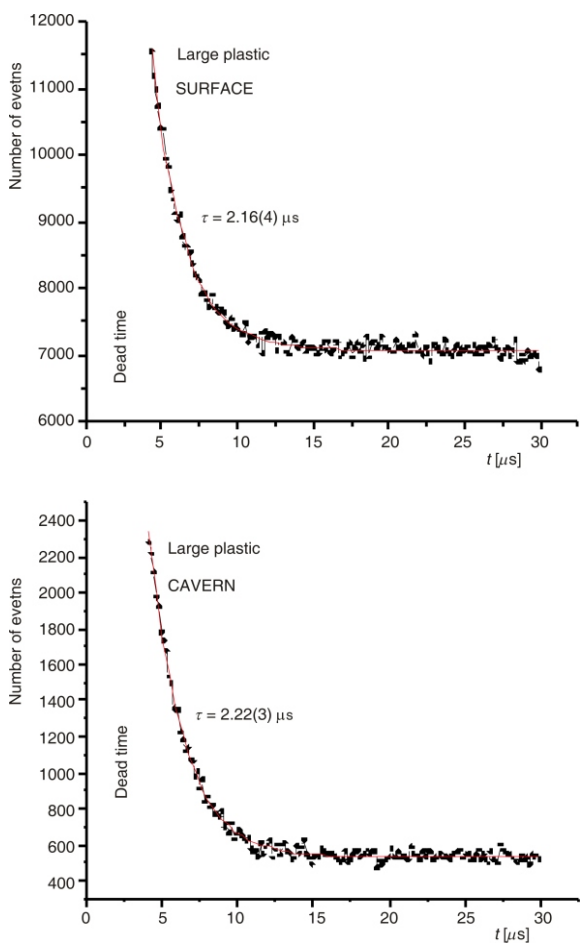

**Figure 13. The distributions of short time intervals between successive counts in large plastic scintillators which sit on the distribution of time intervals between successive counts with a long exponential constant that corresponds to the total counting rate and appears flat on this scale. The decay constant of short time interval distributions, however, equals that of a muon mean life. Note that the first two lifetimes are missing, due to the dead time of the system, which in this particular case equals 4 μs**

It is again exponential, now with the time constant of 2.16(4) and 2.22(3) $\mu$s in the GLL and UL, respectively, reproducing satisfactorily the muon mean life. This suggests that these events originate from muons that both stop and decay within the detector. Minding that the fiducial volume for this kind of signature has not been estimated, the intensity of this exponential distribution, taking in account the missing events due to the 4 s long dead time, now gives the estimate for the lower limit of the number of muons that stop in 5 cm of plastic per square meter per second, at ground level as $6 \cdot 10^{-2}$ $\mu_{stop}$ per m²s, and at a depth of 25 m.w.e. as $1.52 \cdot 10^{-2}$ $\mu_{stop}$ per m²s. It is interesting to compare those figures with the results obtained recently at Gran Sasso [18, 19].

## DECOHERENCE CURVES AND SPECTRA

In test measurements, coincidence spectra between large and small detectors at different separations between the two were recovered in off-line analyses, both in the GLL and the UL. These are predominantly the spectra of EM showers, as seen by respective detectors. The comparison with their direct spectra, which at lower energies are composed mostly of the signatures of environmental gamma radiations and at higher energies of the signatures of CR muons,

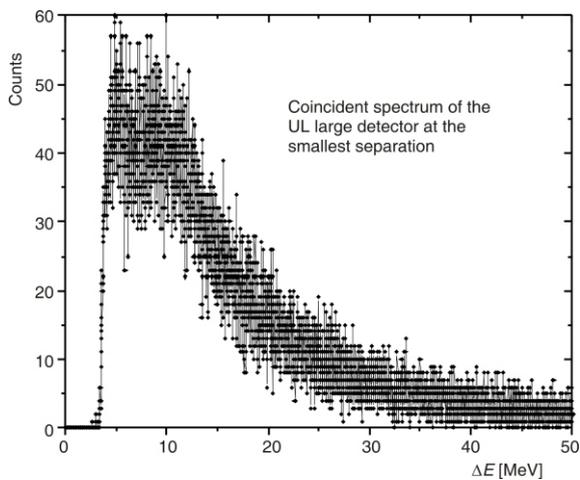

**Figure 14. The coincident spectrum of the large plastic detector in the UL, with the small detector at the smallest separation between them. Compare this with the direct spectrum presented in fig. 9**

enables the disentanglement of the signatures of these radiations. As an illustration, fig. 14 presents the spectrum of the large plastic detector in the UL, in coincidence with the small detector, at their smallest possible separation. This is to be compared with the direct spectrum of the large detector, as presented in fig. 9. Reflecting the structure of the EM component at the given location, not only the intensity, but also the shape of the spectrum changes with detector separation, and the differences between these coincidence spectra in the GLL and the UL reflect the difference in the composition of showers on the surface and underground. The simplest integral characteristic of the shower profile is presented by the integral of this coincidence spectrum as a function of detector separation, sometimes referred to as the decoherence curve. Figure 15 shows the decoherence curve at the GLL, fig. 16 the same in the UL.

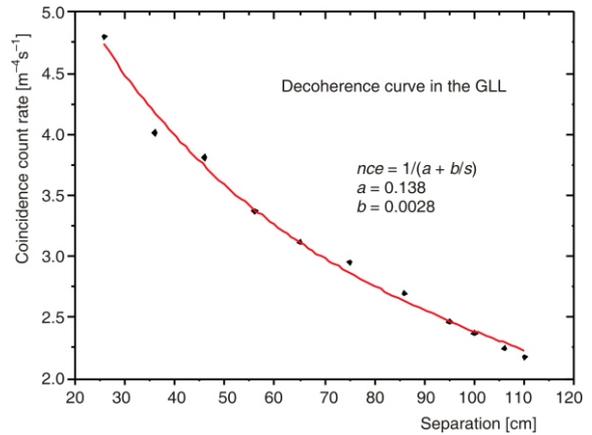

**Figure 15. The decoherence curve in the GLL, reflecting the lateral profile of EM showers on the surface; the nce denotes number of coincident events**

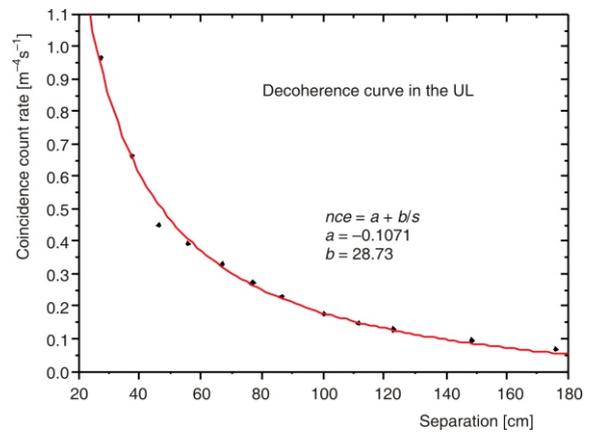

**Figure 16. The decoherence curve in the UL, reflecting the lateral profile of EM showers at the equivalent depth of 25 m.w.e; the nce denotes number of coincident events**

Note that the two curves cannot be satisfactorily fitted with quite the same type of functional dependence, so that the width of the distributions on the surface and underground cannot be directly compared. The much narrower distribution underground is a result of the harder CR muon spectrum and of the different radiation and attenuation lengths, as well as of the geometry of the shower-producing medium. The ratio of the intensities of the distributions on the surface and underground is roughly twice the ratio of CR muon intensities at two locations. Full interpretation awaits better statistics.

## CONCLUSION

We have presented some preliminary results for muon and neutron fluxes at ground level and underground spaces of the Belgrade low-level and CR laboratory, obtained during the commissioning of the new equipment consisting of two scintillation detectors, a HPGe detector, and a digital spectroscopy system based on two CAEN N1728B units. The main advantage of the present set-up is that it enables complex measurements involving routine low activity measurements with modest means, along with some interesting research work related to cosmic-ray physics. We find that the results obtained in this testing phase justify the planned program of measurements and that the future improvement on statistics will contribute not only to the quality of the results already obtained, but furthemore with increased sensitivity new results are also expected to emerge.

## ACKNOWLEDGEMENT

The present work was funded by the Ministry of Education and Science of the Republic of Serbia, under the Project No. 171002. The Belgrade Laboratory bears the name of "Dr. Radovan Antanasijević", in honor of its early deceased founder and first director.

Александар ДРАГИЋ, Владимир И. УДОВИЧИЋ, Радомир БАЊАНАЦ,  
Дејан ЈОКОВИЋ, Димитрије МАЛЕТИЋ, Никола ВЕСЕЛИНОВИЋ,  
Михаило САВИЋ, Јован ПУЗОВИЋ, Иван В. АНИЧИН

## НОВА ОПРЕМА У БЕОГРАДСКОЈ ЛАБОРАТОРИЈИ ЗА МЕРЕЊЕ НИСКИХ АКТИВНОСТИ И КОСМИЧКОГ ЗРАЧЕЊА

Београдска лабораторија састоји се од два лабораторијска простора, једног на површини и једног подземног, на дубини од 25 метара воденог еквивалента. Детаљно су описани и илустровани потенцијали ових лабораторија за мерење ниских активности и за континуирано мерење мионске и електромагнетне компоненте космичког зрачења, као и за студије процеса које ова зрачења индукују у германијумским спектрометрима смештеним у нискофонским подземним лабораторијама. Сва ова мерења се изводе симултано, новим системом за дигиталну спектроскопију, а подаци се записују догађај по догађај, и анализирају после завршених мерења. Такође су приказани прелиминарни резултати који су у фази тестирања опреме добијени за флукс брзих неутрона и заустављених миона у површинској и у подземној лабораторији.

*Кључне речи: подземна лабораторија, гама спектроскопија, мерења ниских активности, космичко зрачење*